\documentclass[twocolumn,showpacs,preprintnumbers,amsmath,amssymb]{revtex4}
\usepackage{graphicx}% Include figure files
\usepackage{dcolumn}% Align table columns on decimal point
\usepackage{bm}% bold math

\begin{document}

%\preprint{}

\title{
Description of the Chemical Reaction Path in the HCO Molecule: \\
A Combined Configuration Interaction and Tight-Binding Approach }

\author{
N.C. Bacalis and A. Metropoulos
}

\address{
Theoretical and Physical Chemistry Institute, National Hellenic
Research Foundation, \\ Vasileos Constantinou 48, GR - 116 35
ATHENS, Greece }

\author{
D.A. Papaconstantopoulos }

\address{
Center for Computational Materials Science, Naval Research
Laboratory, Washington DC 20375-5345 USA }

\date{\today}

\begin{abstract}
% Abstract

It is demonstrated that the reaction path for a polyatomic
molecule (applied to the HCO molecule) is easily calculated via
geometry-independent tight binding Hamiltonian fitted to accurate
\textit{ab-initio} configuration interaction (CI) total energies.
This Hamiltonian not only reproduces the CI calculations
accurately and efficiently, but also effectively corrects any CI
energies happening to erroneously converge to excited states.

\end{abstract}

%3-4 keywords here

\pacs{
31.10.+z %Theory of electronic structure, electronic transitions, and chemical binding in atoms and molecules
31.15.-p %Calculations and mathematical techniques in atomic and molecular physics excluding electron correlation calculations
31.50.-x %Potential energy surfaces (atoms and molecules)
82.20.Kh  %Potential energy surfaces for chemical reactions
}

Further Correspondence: N.C. Bacalis, Fax: +30(210)7273794,
E-mail: nbacalis@eie.gr

\maketitle

% body of paper here

\emph{The question.}
The determination of the reaction path in a chemical reaction
needs the detailed knowledge of the pertinent potential energy
surface (PES) (diabatic or adiabatic). This is a formidable task
because (i) the PES is a multi-dimensional surface, impossible to
be \textit{ab-initio} calculated at every point in the degrees of
freedom (DOF) space, and interpolation is necessary (the most
accurate known detailed multi-dimensional PES is that of H$_3$
interpolated from 71969 \textit{ab-initio} DOF points
\cite{RH372k}). (ii) Because the standard \textit{ab-initio}
calculations of the many electron problem in the Born-Oppenheimer
approximation, based on the variational principle, (accurate
\textit{ab-initio} configuration interaction (CI) calculations)
yield only adiabatic curves, and, more importantly, being
iterative, sometimes converge to undesirable states
(\cite{Rmcscffail}, also c.f. below). Yet, such calculations may
be inhibitively time consuming. For the ground state, the time
problem is already traditionally overcome via the density
functional theory (DFT) \cite{Rdft}, which self-consistently
approximates the many electron by a one-electron problem. However,
DFT calculations sometimes fail to explain experimentally observed
features of the PES \cite{Rdftfail}. Thus, the accurate CI
calculations are more or less indispensable, even if performed in
a rather limited, but representative, set of molecular geometries.
Therefore, a reliable interpolation scheme for the pertinent PES,
based on CI calculations, and overcoming the problem of wrong CI
convergence, is desirable.

\emph{The purpose.}
It is shown that such an interpolation scheme is possible, based
on a spin-polarized \cite{RNRLncb} geometry-independent
\cite{RNRL} Slater - Koster (SK) parametrization \cite{RSK} of
\textit{ab-initio} CI total energies \cite{RMOLPRO}. As a
demonstration, the method is applied to a triatomic molecule of
chemical kinetics interest, HCO, which is an intermediate radical
in the generation of a primary ion during hydrocarbon combustion:
\begin{eqnarray}\label{reaction}
O (^3P) + CH (X^2 \Pi ~\textrm{and/or}~ a ^4\Sigma^-)
& \rightarrow & [HCO]^* (^2A') \nonumber \\
& \rightarrow & HCO^+ + e^- \nonumber
\end{eqnarray}
The reaction of O($^3$P) with CH(X$^2\Pi$, a$^4\Sigma^-$) is known
experimentally \cite{RVGB} to generate the HCO$^+$ cation via
autoionization of some state (or states) of the intermediate HCO
radical upon interaction with some vibrational level of the ion.
The first step toward computations of such interactions is the
construction of the potential energy surface (PES) of the states
with low (or no) barrier, through which a reaction at the
experimental temperature can proceed. Such a state (without a
barrier) is the HCO(X$^2$A') state \cite{Rametamav}, used here to
test the interpolation scheme on a molecular system. The reaction
path of the formation of HCO (in $^2A'$ symmetry) is also computed
using the interpolated PES.

\emph{The procedure.} First several (724 - compared to 71969 of
H$_3$ \cite{RH372k}) accurate CI total energies, based on (less
accurate) multi-configuration self-consistent field (MCSCF)
orbitals, are calculated at selected geometries of the H,C,O atoms
in the A$'$ symmetry of the Cs group. Most of them (508) are
fitted to the interpolation scheme, the remaining serving to check
the quality of the fit. For the fit a non-orthogonal
spin-polarized tight binding (TB) Hamiltonian is formed, whose
matrix elements, along with those of the overlap matrix, are
expressed as functions of the bond direction, according to the SK
scheme \cite{RSK}, and of the bond length, according to the Naval
Research Laboratory (NRL) technique \cite{RNRL}, i.e.: The
functions are generally polynomials of the interatomic distance,
within exponential envelopes, the coefficients and the exponents
being varied as parameters. For two adiabatic states near some
(avoided) crossing the TB Hamiltonian naturally produces two
diabatic PESs in nearby extrapolation, and predicts to which
diabatic PES, ground-state or excited, nearby CI energies belong.
Among these, the appropriate ones can be used to extend the fit
beyond the (avoided) crossings, around which two sets of
parameters are needed for the two PES's. If it happens, as with
HCO, that the ground and excited state energies beyond the
crossing lie close to each other, the adiabatic PES can be fitted
as well, with comparable accuracy.

Finally, by using at each point of the DOF space the lowest lying
TB-fitted PES, the adiabatic path can be found: For each value of
a desired degree of freedom (in our case for each C-O distance)
the energy minimum is searched \cite{Rgram} in the space of the
remaining degrees of freedom (C-H distance and H-C-O angle).
Having the parametrized tight binding Hamiltonian, any property
can be trivially computed.

\emph{Methodology.} For the CI energies the correlation consistent
aug-cc-pVTZ basis set was used \cite{RDunning, RWilson} in
conjunction with the complete active space self-consistent field
(CASSCF) + 1 + 2 multi-reference CI method (MRCI) employed in the
MOLPRO package \cite{RMOLPRO} (the four electrons in the 1s
orbitals of C and O were unexcited). The CASSCF calculations were
state-averaged, and the active space was limited to the 9 valence
orbitals among which the remaining 11 electrons were distributed.
In the subsequent MRCI calculations the uncontracted
configurations were around 50 million internally contracted to
about one million. Calculations between C-O distances of 1.7 and 6
bohr were done for several H-C-O angles between 50$^o$ and 180$^o$
and several C-H distances between 1.7 and 4.5 bohr, most around
the C-H equilibrium distance of 2.12 bohr. The three lowest roots
of the secular equation were computed to increase the accuracy of
the calculation. By an analytic gradient optimization at the MCSCF
level, an approximate (MCSCF) equilibrium geometry was found at
the DOF space point ($\tilde{r}_{HC}, \tilde{r}_{CO},
\tilde{\theta}_{H-C-O}$) = (2.12, 2.2, 126$^o$) (in a.u.). Because
it is not evident whether the aforementioned points are beyond any
avoided crossing, where the role of the ground and the excited
states would be interchanged, first several DOF points near
equilibrium were obtained by employing a generalization of the
3-dimensional sphere to the generally multi-dimensional (in this
case also 3-dimensional) DOF space: $x_i = r_i/\tilde{r}_i -1$, i
= \{HC, CO\}, $x_3 = \theta/\tilde{\theta}-1$, where generally for
$n$ degrees of freedom, points belonging to a $n$-dimensional
hypersphere of radius $r$ and center ($\tilde{x}_i$, i = 1,...,n)
are obtained by
\begin{eqnarray}
x_n - \tilde{x}_n &=& r~cos\theta_n \nonumber \\
x_{n-1} - \tilde{x}_{n-1} &=& r~sin\theta_n~cos\theta_{n-1} \\
... \nonumber \\
x_1 - \tilde{x}_1 &=& r~sin\theta_n~sin\theta_{n-1} ...
cos\theta_1 \nonumber
\end{eqnarray}
where the 1st $\theta_1 =$ 0 or 180$^o$, the two points of a
``1-dimensional sphere", and the other $ 0<\theta_i<180^o$ are the
``azimuthal" hypersphere angles (incidentally, a \emph{variable
dimensional} do-loop code was invented, needed to treat any larger
molecule). Thus, first points with small $r$ were fitted, and
gradually the fit was extended to more remote DOF points.

The formalism of the NRL geometry - independent TB parametrization
is described in detail in Ref. \cite{RNRL}; here an essential
summary is only presented. The total energy is written as
\begin{eqnarray}\label{totalE}
%$
  E[n(\vec{r})] &=& \sum_{i~;~ s=1,2} f(
  \frac{
  \mu - \epsilon_{i~s}
}{
  T
} )
 ~\epsilon_{i~s}
  + F[n(\vec{r})] \nonumber \\
  &\equiv& \sum_{i~;~ s=1,2} f(
  \frac{
  \mu' - \acute{\epsilon_{i~s}}
}{
  T
} )
  ~\acute{\epsilon_{i~s}}
%$
\end{eqnarray}
where \cite{RGillan}
%\begin{equation}\label{fx}
$
  f(x) = 1/(1+e^{x}),
$
%\end{equation}
T=0.005 mRy, and
\begin{equation}\label{eprime}
%$
 \acute{\epsilon_{i~s}} = \epsilon_{i~s} + V_{0} ~~;~~
 \mu' = \mu +  V_{0} ~~;~~
 V_{0} = F[n(\vec{r})]/N_{e}
%$
\end{equation}
with
%\begin{equation}\label{Ne}
$
 N_{e} = \sum_{i~;~ s=1,2} f(
% \frac{
 (\mu -\epsilon_{i~s})/T)
%}{
% T
%} )
$
%\end{equation}
being the number of electrons, $i$ counts the states, $s=1,2$
counts the spin. Since the total energy is independent of the
choice of zero of the potential, the shift $V_0$ is sufficient to
be determined by the requirement that $\acute{\epsilon_{i~s}}$ are
the eigenvalues of the generalized eigenvalue problem $( {\bf
H-S}~\acute{\epsilon_{i~s}} )~\psi_{i~s} =0$, where {\bf H} is the
TB Hamiltonian and {\bf S} is overlap matrix in an atomic s- and
p-orbital basis representation $\{\phi_a\}$. Thus, a
non-orthogonal TB calculation uses on-site, hopping and overlap
parameters. Demanding that only the on-site SK parameters are
affected by the shift $V_{0}$, for atom $I$ in a spin-polarized
structure the matrix elements are expressed as
\begin{equation}\label{hils}
%$
 h^{I}_{l~s} = \sum_{n=0}^{3} b^{I}_{l~n~s}~~\varrho_{I~s}^{2n/3}
 ~~;~~
 l=s,p %,d_{e_{g}},d_{t_{2g}}
%$
\end{equation}
where
\begin{equation}\label{ro}
%$
 \varrho_{I~s}=\sum_{J\neq I}
 e^{-\lambda_{\tilde{I}~\tilde{J}~s}^{2} R_{I~J}}
 ~f(
 \frac{
   R_{I~J}-R_{0}
}{
   r_{c}
} )
%$
\end{equation}
is a generalized pair potential (``density"), with  $R_{0}$ = 15
bohr, $r_{c}$ = 0.5 bohr, $R_{I~J}$ is the internuclear distance
between atoms $I$ and $J$, $\tilde{I}(\tilde{J})$ denote the type
of atom on the site $I(J)$ while
$\lambda_{\tilde{I}~\tilde{J}~s}$, depending on the atom type, and
$b^{I}_{l~n~s}$ are the on-site NRL geometry-independent
parameters (GIP). It is found sufficient to keep hopping and
overlap parameters spin independent, of the form
\begin{equation}\label{Pgamma}
%$
 P_{\gamma}(R)=
 (\sum_{n=0}^{2} c_{\gamma~n}~R^{n})~e^{-g_{\gamma}^{2}R}
 ~f(
 \frac{
   R-R_{0}
}{
   r_{c}
} )
%$
\end{equation}
where $\gamma$ indicates the type of interaction (i.e. $ss\sigma$,
$sp\sigma$, $pp\sigma$, $pp\pi$ and $ps\sigma$). The NRL GIPs are
$c_{\gamma~n}$ and $g_\gamma$~, $R$ is the interatomic distance,
and $R_{0}$ and $r_{c}$ are as in eq. \ref{ro}.
%Here we do not distinguish between $t_{2g}$ and $e_{g}$.

%Although the standard NRL parameters allow for s-, p- , and
%d-matrix elements, in this case s- an p-matrix elements were
%proven sufficient.
Within the context of the NRL code \cite{RNRL}, written primarily
for solids, the molecule was treated as a base to a large cubic
lattice unit cell (lattice constant = 100 a.u.) ensuring vanishing
interaction between atoms in neighboring cells. Thus, the PES was
described in terms of the following NRL GIPs for each spin
polarization. On-site: s: H, C, O, (H depending on C), (C on H),
(H on O), (O on H), (C on O), and (O on C); p: C, O, (C on H), (O
on H), (C on O), and (O on C). Hopping and overlap parameters:
ss$\sigma$: H-C, H-O, C-O; sp$\sigma$: H-C, H-O, C-O and O-C
(denoted as ps$\sigma$); pp$\sigma$ and pp$\pi$: C-O. For HCO,
since similar atoms are well separated, the H-H, C-C and O-O
parameters vanish. We fitted 508 CI points and checked the
resulting PES against 216 more CI energies not included in the
fit. The error was less than 10$^{-3}$ a.u., which is within the
\textit{ab-initio} PES accuracy (starting from different initial
guesses the MCSCF calculation may converge to slightly different
results by 10$^{-3}$ a.u.). To ensure obtaining physically
meaningful TB parameters, for a very limited number of molecular
geometries the Hamiltonian eigenvalues were also fitted, while the
total energy was fitted for all 508 structures.

Finally, for the reaction path we used a non-linear energy
minimization technique employing Powell's conjugate directions
method \cite{Rnumrec} modified to be restricted to closed
intervals of the DOF space \cite{Rgram}.

For comparison, each of the 724 \textit{ab-initio} CI calculations
needs 3 hours of CPU time, each $n$-dimensional hypersphere radius
$r$-increase, to fit more remote points (with 10 such hypersphere
radial extensions all points can be covered) needs 2-3 hours, and
each 2-dimensional energy minimization, using the final TB
parameters (i.e. the reaction path determination), needs a few
seconds.

\begin{figure}
\includegraphics[width=3.5in]{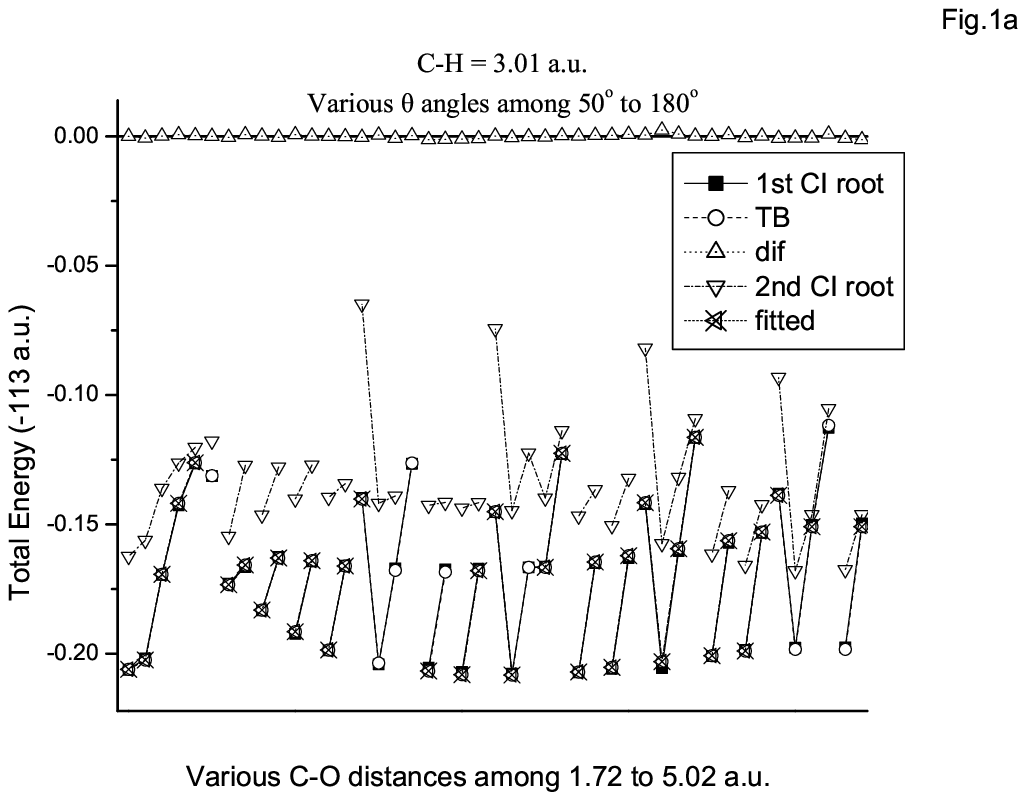}% Here is how to import EPS art

\includegraphics[width=3.5in]{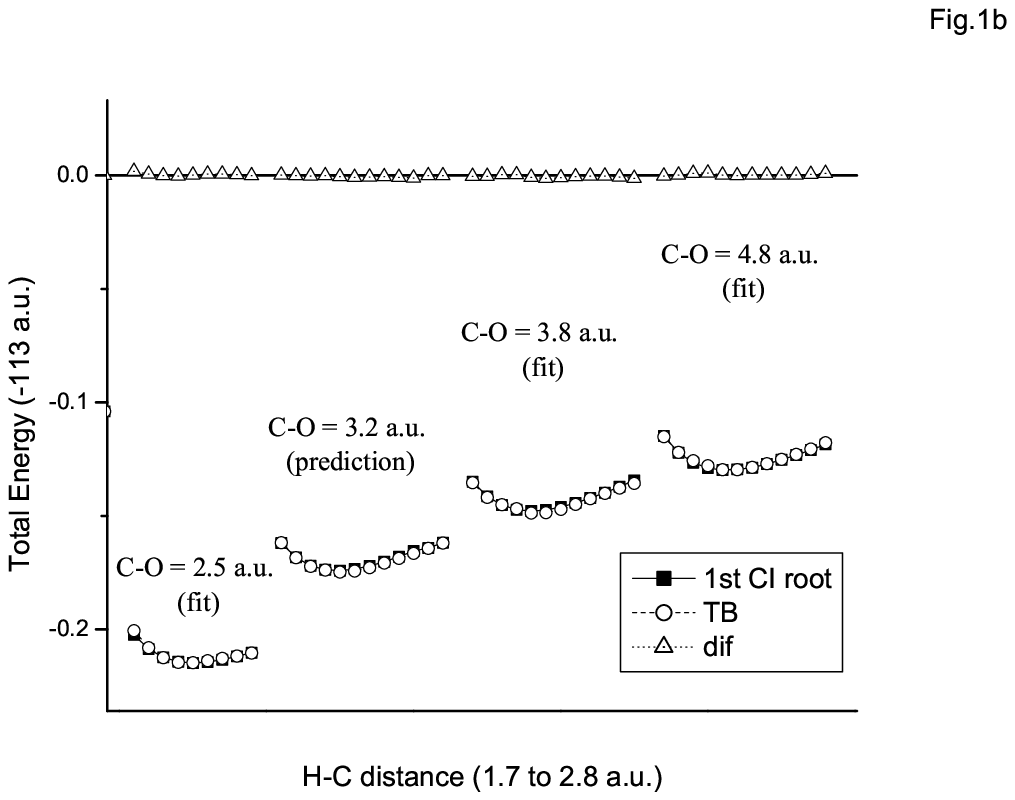}% Here is how to import EPS art
\caption[]{ Predicted total energy E in a.u. (Above:) vs C-O
distance for C-H distance = 3.01 bohr, and various H-C-O angles.
(Below:) vs C-H distance for various C-O distances, and H-C-O
angle = 100$^o$.
}\label{predict}
\end{figure}

\begin{figure}
\includegraphics[width=3.5in]{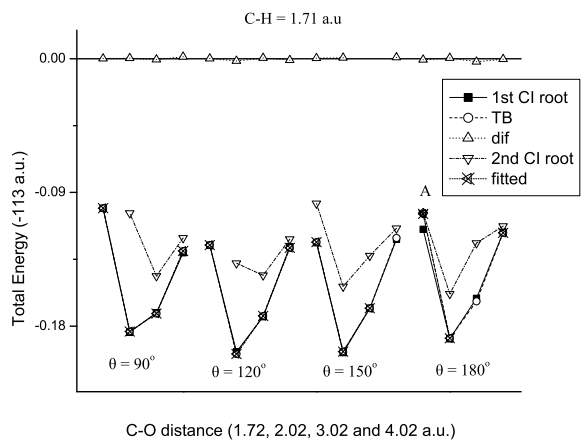}% Here is how to import EPS art
\caption[]{ The CI point A (excited) in E vs C-O distance for C-H
distance = 1.71 bohr and H-C-O angle =% 150$^o$ and
180$^o$ is predicted by the fit to belong to the diabatic branch
of the curve beyond the avoided crossing. (Inclusion of the lower
value to the fit, destroys it.)
} \label{cross}
\end{figure}

\begin{figure}
\includegraphics[width=3.5in]{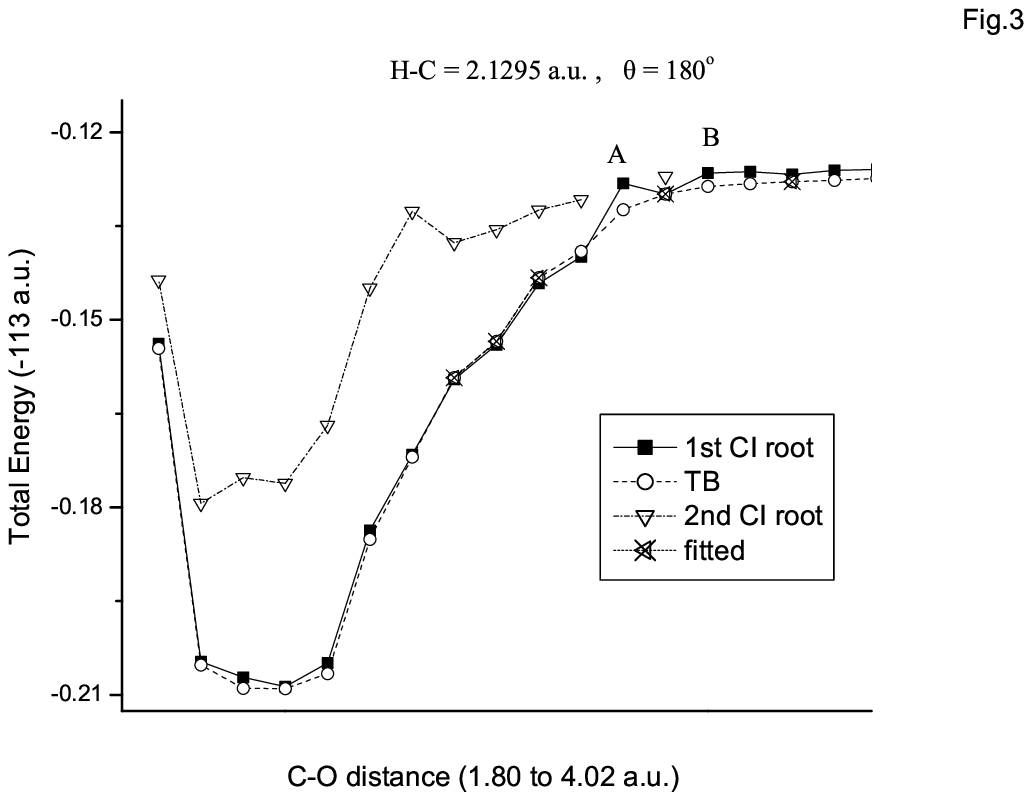}% Here is how to import EPS art
\caption[]{ The CI points A and B clearly belong to the excited
state as shown by the TB prediction. The CI calculation could not
converge to the correct values. The discontinuity can be verified
by observing the corresponding MCSCF orbitals.
} \label{excited}
\end{figure}

\begin{figure}
\includegraphics[width=3.5in]{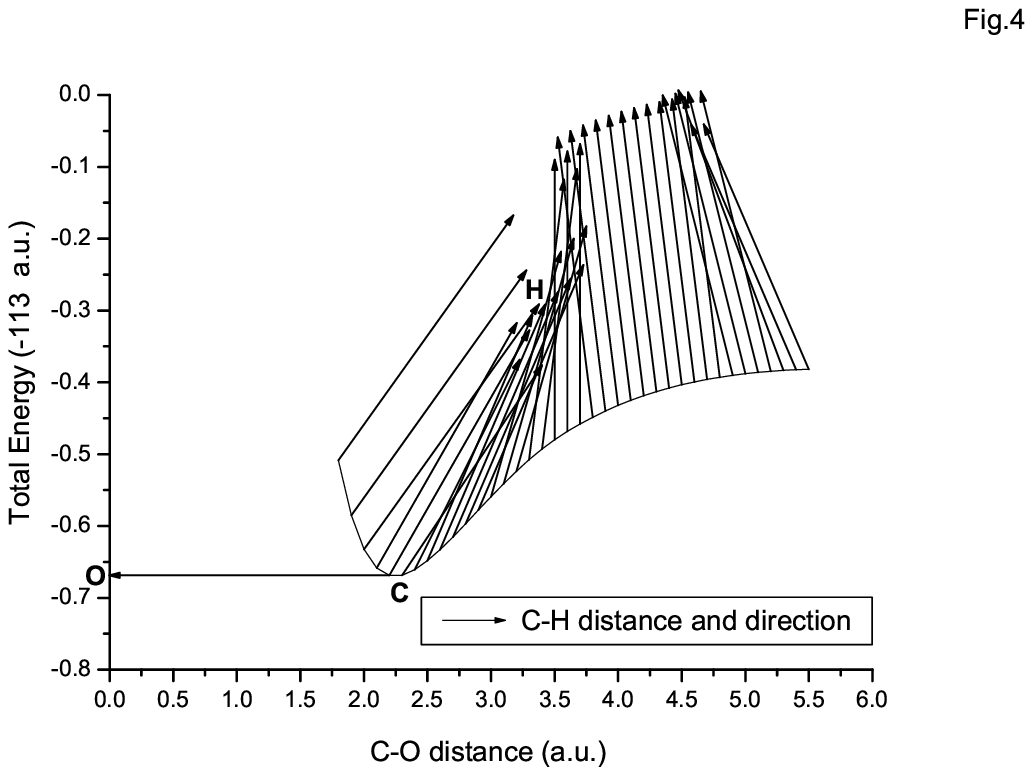}% Here is how to import EPS art
\caption[]{
The reaction path for the formation of HCO. Details
are described in the text.
} \label{reactionpath}
\end{figure}

\emph{Results.}
The fitted TB Hamiltonian could predict correctly total energy
curves for points not included in the fit as shown for example in
Fig. \ref{predict}. Since it produces naturally the diabatically
extended branch of the energy, it could distinguish to which
adiabatic state near an avoided crossing the CI values belong.
Classifying such CI points may sometimes be misleading or
unrecognizable by mere observation of the MCSCF orbitals. An
example is shown in Fig. \ref{cross}. However, the most impressive
aspect was that we realized, through the fit, that at some points
(about 10 in 700) the CI calculation had converged to
\emph{excited} energies (which ought to be disregarded, otherwise
they would destroy the fit). An example is given in Fig.
\ref{excited}. Finally, Fig. \ref{reactionpath} shows the reaction
path for the formation of HCO, as HC approaches O. For a triatomic
molecule the figure contains the whole information: For each C-O
distance the minimum energy and the corresponding C-H distance and
H-C-O angle are displayed. At large C-O distances, O is more
attracted toward H, but, in approaching equilibrium, O binds
mainly with C, the H-C-O angle gradually becoming $\simeq$ 122$^o$
(representing the CI value). Around equilibrium (c.f. Table
\ref{Teq}), the angle changes slightly monotonically by 1-2$^o$,
but because, in increasing the C-O distance, the C-H distance
decreases, predominantly an antisymmetric stretching vibration
occurs. To our knowledge there is no experimental confirmation of
the reaction path of this intermediate molecule.

\emph{Acknowledgment:}
We wish to thank Dr. M.J. Mehl for many useful discussions.

%\newpage
%


% Fig.1
% \narrowtext

% Fig.2

% Fig.3

% Fig.4

\begin{table}
\caption[]{Geometric characteristics of HCO around equilibrium,
along the reaction path, in a.u. (H-C-O angle in degrees). The
last three columns indicate the minimum energy molecular geometry.}
\label{Teq}
\begin{ruledtabular}
\begin{tabular}{cccc}
C-O distance & Total Energy & C-H distance & H-C-O angle \\
\cline{2-4}
%\colrule
2.6 &  -113.6328  &  2.069 &  117.53 \\
2.5 &  -113.6485  &  2.071 &  118.69 \\
2.4 &  -113.6610  &  2.077 &  119.77 \\
2.3 &  -113.6685  &  2.088 &  120.84 \\
2.2 &  -113.6687  &  2.107 &  121.91 \\
2.1 &  -113.6583  &  2.130 &  122.98 \\
2.0 &  -113.6326  &  2.153 &  124.09 \\
%\hline

\end{tabular}
\end{ruledtabular}
\end{table}

\end{document}